\documentclass[aps,prl, amsmath, showpacs, preprintnumbers,superscriptaddress, twocolumn,sort&compress, amssymb]{revtex4}

\usepackage{graphicx}
\usepackage{mathptmx, textcomp}
\usepackage[latin1]{inputenc}
\begin{document}
\title{Collisions between tunable halo dimers:\\ exploring an elementary four-body process with identical bosons}
\author{F. Ferlaino}
\author{S. Knoop}
\author{M. Mark}
\author{M. Berninger}
\author{H. Sch\"{o}bel}
\author{H.-C. N\"{a}gerl}
\affiliation{Institut f\"ur Experimentalphysik and Zentrum f\"ur Quantenphysik, Universit\"at
 Innsbruck, % Technikerstra{\ss}e 25,
 6020 Innsbruck, Austria}
\author{R. Grimm}
\affiliation{Institut f\"ur Experimentalphysik and Zentrum f\"ur Quantenphysik, Universit\"at
 Innsbruck, % Technikerstra{\ss}e 25,
 6020 Innsbruck, Austria}
\affiliation{Institut f\"ur Quantenoptik und Quanteninformation,
 \"Osterreichische Akademie der Wissenschaften, 6020 Innsbruck,
 Austria}

\date{\today}


\begin{abstract}
We study inelastic collisions in a pure, trapped sample of Feshbach molecules
made of bosonic cesium atoms in the quantum halo regime. We measure the
relaxation rate coefficient for decay to lower-lying molecular states and study
the dependence on scattering length and temperature. We identify a pronounced
loss minimum with varying scattering length along with a further suppression of
loss with decreasing temperature. Our observations provide insight into the
physics of a few-body quantum system that consists of four identical bosons at
large values of the two-body scattering length.

\end{abstract}

\maketitle

Few-body quantum systems in halo states exhibit unique properties
\cite{Jensen2004sar}. A quantum halo is a very weakly bound state whose wave
function extends far into the classically forbidden range. Halo systems are
much larger than one would expect from the characteristic interaction range of
their constituents. Many examples for halo states are known in nuclear physics,
with the deuteron being a prominent example \cite{Blatt1952book}. In molecular
physics, the He dimer has for many years served as the prime example of a halo
state \cite{Luo1993, Schollkopf1994}.

A particular motivation to study halo states is given by the concept of
universality in few-body systems \cite{Braaten2006uif}. Since, in a halo state,
short-range details of the interaction become irrelevant, the system is
described by very few parameters and shows universal behavior in its low-energy
observables. A halo dimer is the elementary two-body halo system. Here the only
relevant length scale is given by the scattering length $a$, which describes
the $s$-wave interaction between its two constituents. The size of the halo
dimer is directly related to $a$ and the binding energy is $E_b = \hbar^2/(2
\mu a^2)$, where $\mu$ is the reduced mass. For three-body halo systems,
universal Efimov states can exist \cite{Efimov1970ela,Kraemer2006efe}. Here one
additional parameter is required to fully describe the system in the universal
limit. A natural further step is to investigate the universal physics of
systems composed of four identical bosons. The fundamental properties of such
systems are unexplored terrain, with the existence of an additional four-body
parameter \cite{Yamashita2006fbs,Hammer2007upo}, the scaling and threshold
behavior, and the binding energies of four body-states \cite{Hammer2007upo}
being open issues.

In the field of ultracold gases, the Feshbach association technique
\cite{Kohler2006poc} has provided experimentalists with unprecedented
possibilities to create and study halo dimers. In the case of certain Feshbach
resonances, a considerable range of universality exists, where halo dimers can
be conveniently controlled by a magnetic bias field to vary their binding
energy and size. Such \textit{tunable halo dimers} are unique probes to explore
quantum phenomena related to universality. A binary collision between two halo
dimers can be seen as an \textit{elementary four-body process}. For a special
kind of halo dimers such a four-body process has already attracted considerable
attention: halo dimers made of fermionic atoms in different spin states allow
to create molecular Bose-Einstein condensates and to study the crossover to a
fermionic superfluid \cite{Inguscio2006ufg}. Here a key point is the Pauli
suppression effect that, in combination with the halo nature of the dimer,
leads to stability against decay into lower-lying molecular states and favors
elastic processes \cite{Petrov2004wbm}.

In this Letter, we study binary collisions in a pure, trapped sample of tunable
halo dimers made of \textit{bosonic} atoms. Halo dimers of this class have so
far received much less experimental attention than their fermionic
counterparts, although they represent an important link to universal few-body
phenomena in systems of few interacting bosons; for three particles, an early
example is the predicted atom-dimer ``Efimov'' resonances \cite{Efimov1979lep}.
Halo dimers of bosonic atoms have been realized in ultracold gases of $^{85}$Rb
and $^{133}$Cs \cite{Thompson2005sdo,Mark2007sou} and  properties of the
individual dimers, like binding energies, magnetic moments, and spontaneous
decay rates, have been measured. In contrast, their collision properties have
remained unexplored terrain. Because of the absence of a Pauli suppression
effect, substantial inelastic decay to lower-lying molecular states can be
expected. The observation of loss  serves as a probe for dimer-dimer
interactions \cite{Jensen2004sar,Kohler2006poc}.

\begin{figure}
 \includegraphics[width=8.0cm] {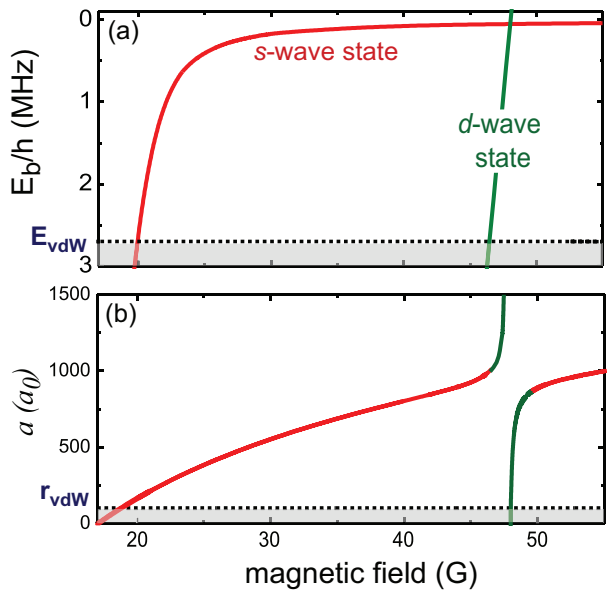}
 \caption{(color online) Weakly bound Cs$_2$ molecular states at low magnetic fields.
 (a) Binding energies of the relevant Cs$_2$ dimer states.
 The zero-energy level is the $s$-wave threshold of two colliding Cs
 atoms in their absolute hyperfine ground state sub-level.
 % with zero kinetic energy.
 (b) $s$-wave scattering length $a$ versus the magnetic field (see text).
 The shaded regions in (a) and (b) indicate the non-halo regime with
 $E_b> E_{\text{vdW}}$, and $a<r_{\text{vdW}}$, respectively. }
 %We populate the
% $s$-wave state by using the near-threshold avoided crossing
% with the $d$-wave state \cite{Mark2007sou}. Higher partial-wave states
% \cite{Chin2004pfs, Knoop2008mfm} are not shown as they are irrelevant
% for the present experiments.}
 \label{fig1}
\end{figure}

Our experiments are performed with $^{133}$Cs atoms, which represent an
excellent system to study few-body physics with bosons at large scattering
lengths \cite{Kraemer2006efe} because of the unique scattering properties
\cite{Chin2004pfs}. In the lowest spin state at low magnetic fields, one finds
a broad entrance-channel dominated $s$-wave Feshbach resonance
\cite{Weber2003bec, Kohler2006poc} along with an extraordinarily large
background scattering length. The scattering properties of ultracold atoms are
governed by the last bound $s$-wave state below the dissociation threshold as
displayed in Fig.\,\ref{fig1}. In a wide magnetic-field range, this state
carries a quantum halo character, where the two-body scattering length far
exceeds the classical interaction range of the van der Waals potential,
$r_{\text{vdW}}$ ($\approx 100a_0$), and the binding energy $E_b$ is much
smaller than the corresponding $E_{\text{vdW}}$ ($\approx h\times 2.7$ MHz)
\cite{vdW}. Here $a_0$ is Bohr's radius.
%The Bohr radius $a_0$ is about 52.9 $10^{12}$ m.} .
%For Cs, $C_6= 6860$~a.\,u.}
%r$_{\text{vdW}}\approx 100a_0$, and E$_{\text{vdW}}\approx h\times
%2.7$ MHz \cite{Chin2004pfs}.}.

Our experimental procedure to produce an optically trapped sample of tunable
halo dimers involves several stages. We initially prepare ultracold trapped
$^{133}$Cs atoms in their absolute hyperfine ground state sublevel $\vert
F,m_{F}\rangle=\vert 3, 3\rangle$, similarly to Ref.\,\cite{Mark2007sou}. The
atoms are optically trapped by two crossed 1064-nm laser beams with waists of
about 250 $\mu$m and 36 $\mu$m, while a magnetic field levitates the atoms
against gravity \cite{Weber2003bec, Herbig2003poa}. The levitation field
ensures an optimized evaporative cooling of the atoms, which is realized by
lowering the optical power in the trapping beams. We stop the cooling just
before the onset of Bose-Einstein condensation to avoid   too high atomic
densities. At our lowest temperature of 20~nK, we obtain about $1.5\times10^5$
non-condensed atoms.

In the next stage, we create the halo dimers by Feshbach association. Here, the
application of the levitation field is not appropriate since atoms and dimers
in general have different magnetic moments. This leads to special requirements
for the trap design. On the one hand, we need a sufficiently high optical
gradient in the vertical direction to hold the atoms and dimers against
gravity. On the other hand, we want to avoid the high density of a tight trap,
which causes fast losses driven by atom-dimer collisions. These two
requirements can be simultaneously fulfilled by using an elliptic trap
potential with weak horizontal confinement and tight confinement in the
vertical direction. Shortly before molecule production, we adiabatically
convert the levitated trap to a non-levitated trap by simultaneously changing
the levitation field, the optical power, and the trap ellipticity. The latter
is modified by a rapid spatial oscillation of the 36-$\mu$m waist beam in the
horizontal plane with the use of an acousto-optic modulator at a frequency of
about 100~kHz, which greatly exceeds the typical trap frequencies, and thus
creates a time-averaged optical potential \cite{Milner2001obf, Friedman2001ooc,
AltmeyerPRA2007}. The adiabatic change of the trap shape is set in a way to
keep the peak density, the temperature, and thus the phase-space density
constant \cite{trapfreq}. Right after converting the trap, we associate the
halo dimers by sweeping the magnetic field across the 200~mG wide $d$-wave
Feshbach resonance located at approximately 48 G \cite{Mark2007sou}, (see
Fig.\,\ref{fig1}).

\begin{figure}
 \includegraphics[width=8.0cm] {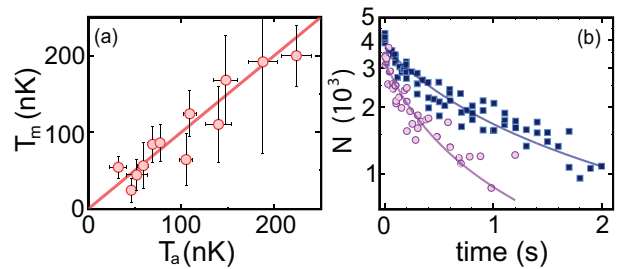}
 \caption{(color online) Measurements on the trapped molecular sample.
 (a) Comparison between the atomic
 and molecular temperatures $T_{\text{a}}$ and $T_{\text{m}}$ at 35G, where $E_\text{b}\simeq h \times 87$~kHz.
 The solid line indicates equal temperatures for atoms and molecules.
 (b) Number of halo dimers $N$ as a function of the holding time in trap
  at 28.3~G ($a = 500 a_0$, squares) and at
 45.6 G ($a = 900 a_0$, circles).
 The solid lines are fits to the data (see text).}
 \label{fig2}
\end{figure}

The molecular temperature can be set by adjusting the temperature of the
initial atomic sample. To selectively measure the atomic and molecular
temperatures, we spatially separate the two components with the Stern-Gerlach
technique, and we perform a subsequent time-of-flight imaging, as described in
Refs.~\cite{Herbig2003poa, Mark2007sou}. In Fig.\,\ref{fig2}(a), we compare the
atomic and the molecular temperatures for a wide range of the initial
temperatures. In spite of some overall heating in the conversion process, we
observe that the dimers and the atoms have the same translational temperature
in the trap. This observation may indicate an elastic interaction between atoms
and dimers on the 10~ms time scale of our preparation sequence and it allows us
to conveniently use the atomic sample to determine the temperature of the
molecular gas.

In the last step of the preparation sequence, we selectively remove the atoms
from the dipole trap. This is done by using a double-resonant purification
scheme, which combines microwave excitation with resonant light \cite{purify},
similarly to Ref.\,\cite{Thalhammer2006llf}. By absorption imaging after the
Stern-Gerlach technique we verify that no atoms are remaining. We do not
observe heating or loss of molecules as induced by the purification sequence.

%\begin{figure}
% \includegraphics[width=8.4cm] {pics/Fig3.eps}
% \caption{(color online) Collisional decay of a pure sample of trapped halo dimers at 28.3~G ($a = xxx a_0$, squares) and at 45.6 G ($a = yyy a_0$, circles).
% The solid lines are fits to the data (see text).}
% \label{fig3}
%\end{figure}

We then use the pure, trapped sample of tunable halo dimers to study binary
collisions. We measure inelastic decay, resulting from the relaxation into more
deeply bound states. In such a process the conversion of internal into kinetic
energy by far exceeds the trap depth and leads to immediate trap loss of all
particles involved. All our experiments are carried out in a regime of very low
temperatures ($k_\text{B} T \ll E_\text{b}$), where the initial kinetic energy
of the colliding dimers is not sufficient to break up the molecules, as
observed for $^6$Li halo dimers in Ref.\,\cite{Jochim2003pgo}. Moreover,
spontaneous dissociation observed for $^{85}$Rb halo dimers
\cite{Thompson2005sdo,Kohler2005sdo} is not possible as there is no
energetically open channel. Other density-independent losses, such as
background collisions or light-induced losses, can also be neglected under our
experimental conditions. We can therefore completely attribute the observed
losses to inelastic dimer-dimer collisions. The decay of the trapped dimer
sample is thus  described by the usual rate equation
$\dot{N}=-\alpha_{\text{rel}}\bar{n} N$. Here $N$ indicates the number of
dimers and $\alpha_{\text{rel}}$ the relaxation rate coefficient. The mean
molecular density $\bar{n}$ is given by $\bar{n}=\left[m \bar{\omega}^2/(2\pi
k_B T)\right]^{3/2}N$ with $m$ being the atomic mass and $\bar{\omega}$
denoting the geometric mean of the trap frequencies.

\begin{figure}
 \includegraphics[width=8.0cm] {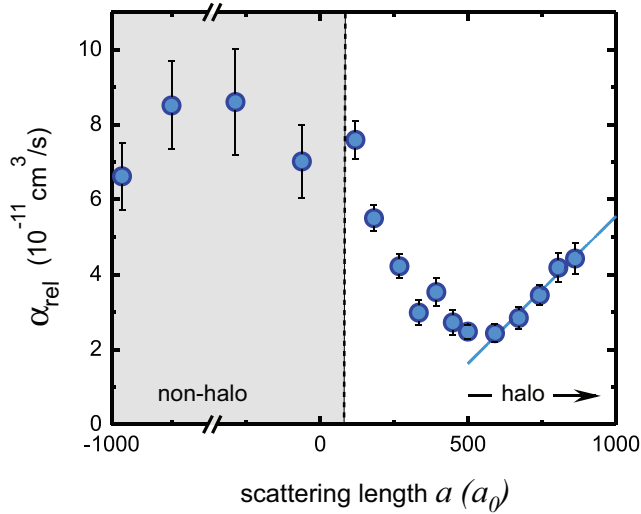}
 \caption{(color online) Scattering length dependence of
 the relaxation rate coefficient $\alpha_{\text{rel}}$ at 120~nK. The solid line is a linear fit to the data
 in the region $a\geq 500 a_0$ (see text). The error bars refer to the statistical uncertainty.
 The shaded region indicates the $a<r_{\text{vdW}}$ regime;
 here the experimental uncertainties are larger because the molecules
 have to be transferred through several avoided crossings with higher
 partial-wave states \cite{Mark2007sou}.}
 \label{fig4}
\end{figure}

We measure $\alpha_{\text{rel}}$ as a function of the  scattering length $a$
for a fixed temperature $T=120$~nK. We ramp the magnetic field to a desired
value, and we then perform a lifetime measurement on the  trapped dimers for
storing times up to 2~s. As an example, the time evolution of the dimer number
in the optical trap is shown in Fig.\,\ref{fig2}(b) at two different values of
the magnetic field. We observe the expected non-exponential decay of the dimer
number. We extract the value of $\alpha_{\text{rel}}$  by fitting the data with
the above rate equation. The lifetime measurements are then repeated at
different values of the magnetic field in a range from 7~G to 50~G.

The observed dependence of the relaxation rate coefficient on the scattering
length reveals an interesting behavior; see Fig.\,\ref{fig4}. Non-halo dimers
exhibit a relative large and essentially constant collisional rate coefficient.
In contrast, when the dimers enter the halo regime with increasing scattering
length ($a>r_{\text{vdW}}$), the rate coefficient first drops to a minimum. The
minimum is found at $a\approx 500 a_0$ ($\sim$5$r_{\text{vdW}}$). For larger
values of $a$, the rate coefficient increases with $a$. General considerations
\cite{Braaten2004,DincaoPRL2006, Petrovprivate} suggest to first approximation
a universal linear scaling law for $\alpha_{\text{rel}}$ according to
$\alpha_{\text{rel}} = C(\hbar/m) a$ with a dimensionless constant $C$. From a
linear fit for $a>500 a_0$ (solid line) we obtain an estimate for $C$ of about
3.

An essentially constant relaxation rate coefficient for varying binding energy
or magnetic field has been previously measured in collisions between non-halo
dimers. Class of the non-halo dimers includes $^{23}$Na$_2$
\cite{Mukaiyama2004dad}, $^{87}$Rb$_2$ \cite{Syassen2006cdo}, $^{133}$Cs$_2$ in
various molecular states \cite{Chin2005oof, Knoop2008mfm}, and also
$^{6}$Li$_2$ $p$-wave molecules as a process involving four identical fermions
\cite{Inada2008cpo}. Our halo dimers composed of bosonic atoms thus show a
novel and qualitatively different behavior.

In a second set of experiments, we study the temperature dependence of the
relaxation rate coefficient. In the ultracold domain, inelastic two-body
collision processes are usually described in terms of a simple rate constant,
i.e.\, a rate coefficient being independent of the particular collision energy.
This applies to the case of inelastic atom-atom collisions \cite{Landau1965},
as well as to collisions between deeply bound dimers \cite{Lee, Quemener2007}.
In contrast to this usual behavior, we find a strong temperature dependence of
the loss rate coefficient of halo dimers, so that a simple rate constant model
does not apply.

\begin{figure}
 \includegraphics[width=8cm] {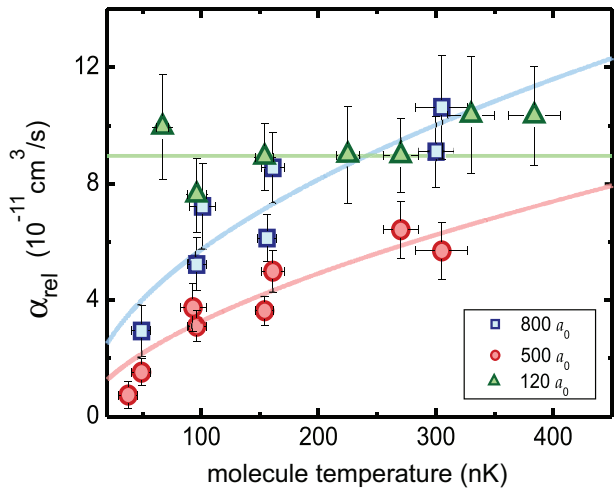}
 \caption{(color online) Temperature dependence of the relaxation rate
 coefficient $\alpha_{\text{rel}}$ at $120~a_0$ (triangles), $500~a_0$ (squares),
 and $800~a_0$ (circles). The solid lines are introduced as guides to the eye.}
 \label{fig5}
\end{figure}

We have focused our measurements on the temperature-dependence in three
different cases: the non-halo regime ($120~a_0$), the loss minimum in the halo
regime ($500~a_0$), and a more extreme halo case with increased loss
($800~a_0$). As shown in Fig.\,\ref{fig5}, the dimers exhibit the expected
constant relaxation rate outside of the halo regime ($\approx 9 \times
10^{-11}$~cm$^3$/s). In the halo regime, we observe a clear decrease of the
relaxation rate with decreasing temperature, both at $500~a_0$ and $800~a_0$,
roughly following a $\sqrt{T}$-dependence. This surprising behavior raises the
question whether the temperature-dependence of the relaxation rate is a
property unique to halo states or whether it can also occur for other weakly
bound Feshbach molecules. In the latter case, a halo dimer may just be seen as
an extreme case of a weakly bound dimer. We speculate that this observation is
related to the fact that collisions between weakly bound dimers can involve
more complex processes which go beyond simple two-body mechanisms. For
instance, the release of binding energy may lead to a fragmentation with three
particles in the exit channel, i.e.\ a more deeply bound molecule and two free
atoms. Breakup thresholds may manifest themselves in a more complicated
dependence on the collision energy.

% RUDI 22 March 2008
The possibility to suppress inelastic loss by controlling temperature and
magnetic field leads to a favorable situation, in which halo dimers exhibit a
high degree of stability.  In particular, at 40~nK and 500~$a_0$, we measure a
loss rate coefficient as low as $\sim 7\times 10^{-12}$~cm$^3$/s. This
unusually small value corresponds to an order of magnitude improvement in the
stability against collisional decay with respect to previously investigated
cases of $^{23}$Na$_2$ \cite{Mukaiyama2004dad} and $^{87}$Rb$_2$
\cite{Syassen2006cdo}.

To conclude, we have studied inelastic collisions between tunable halo dimers
composed of bosonic atoms. We have observed a pronounced scattering length
dependence, with a minimum in the loss coefficient as a most striking feature.
We have also found that inelastic loss is further suppressed by decreasing the
temperature. The existence of the minimum raises the question whether this
feature can be understood in terms of universal four-body physics, similar to a
minimum in three-body recombination at large positive scattering length
\cite{Esry1999rot, Nielsen1999ler, Braaten2006uif, Kraemer2006efe}, which
results from the destructive interference of two decay channels. The slow
inelastic decay near the minimum may provide us with a favorable situation to
study elastic dimer-dimer interactions or to search for universal four-body
bound states \cite{Hammer2007upo}.

We thank B.\,Esry, M.\,Baranov, D.\,Petrov, G.\,Shlyapnikov, and T.\,K\"{o}hler
for fruitful discussions. We acknowledge support by the Austrian Science Fund
(FWF) within SFB 15 (project part 16). S.~K.\,is supported within the Marie
Curie Intra-European Program of the European Commission. F.~F.\,is supported
within the Lise Meitner program of the FWF.

\bibliographystyle{apsrev}

%\bibliography{ultracold,DD}

\begin{thebibliography}{40}
\bibitem[{tra()}]{trapfreq}
\bibinfo{note}{For the lowest temperature samples, the final time-averaged
  elliptic potential has trap frequencies of $10$~Hz and $20$~Hz
  in the horizontal plane, and $80$~Hz in the vertical direction.}
\bibitem[{pur()}]{purify}
\bibinfo{note}{The microwave drives the $\vert 3, 3\rangle \rightarrow \vert 4,
  4\rangle$ hyperfine transition, while the resonant light pulse drives the
  closed optical transition $\vert 4, 4\rangle \rightarrow \vert 5, 5\rangle$,
  which pushes the atoms out of the optical trap. The purification is performed
  at a magnetic field of 35\,G.}
\end{thebibliography}

\end{document}